\documentclass{article}
\usepackage{float}

\def \lket {|}
\def \rket {\rangle}
\def \lbra {\langle}

\def \A {{\cal A}}
\def \B {{\cal B}}
\def \H {{\cal H}}

\newcommand{\ket}[1]{\lket #1\rket}

\newcommand{\comment}[1]{}

\newtheorem{Theorem}{Theorem}
\newtheorem{Lemma}{Lemma}
\newtheorem{Corollary}{Corollary}
\newtheorem{Claim}{Claim}
\newcommand{\proof}{\noindent {\bf Proof: }}
\newcommand{\qed}{\nobreak \ifvmode \relax \else
      \ifdim\lastskip<1.5em \hskip-\lastskip
      \hskip1.5em plus0em minus0.5em \fi \nobreak
      \vrule height0.75em width0.5em depth0.25em\fi}
\floatstyle{boxed}
\newfloat{Algorithm}{hbt}{lop}
\floatname{Algorithm}{Algorithm}

\begin{document}
\title{Variable time amplitude amplification and a faster quantum
algorithm for solving systems of linear equations}
\author{Andris Ambainis\thanks{Faculty of Computing, University of Latvia,
Raina bulv. 19, Riga, LV-1586, Latvia, {\tt ambainis@lu.lv}. 
Supported by ESF project 1DP/1.1.1.2.0/09/APIA/VIAA/044,
FP7 Marie Curie Grant PIRG02-GA-2007-224886 and
FP7 FET-Open project QCS.}}
\date{}

\maketitle

\begin{abstract}
We present two new quantum algorithms. Our first algorithm is a generalization
of amplitude amplification to the case when parts of the quantum algorithm that is being 
amplified stop at different times. 

Our second algorithm uses the first algorithm to improve the running time of 
Harrow et al. algorithm for solving systems of linear equations from $O(\kappa^2 \log N)$ to
$O(\kappa \log^{3} \kappa \log N)$ where $\kappa$ is the condition number of the system
of equations. 
\end{abstract}

\section{Introduction}

Solving large systems of linear equations is a very common problem
in scientific computing, with many applications.
Until recently, it was thought that quantum algorithms
cannot achieve a substantial speedup for this problem,
because the coefficient matrix $A$ is of size $N^2$ and
it may be necessary to access all or most of coefficients in $A$ to
compute $x$ - which requires time $\Omega(N^2)$.
 
Recently, Harrow, Hassidim and Lloyd \cite{HHL} discovered a 
surprising quantum algorithm that allows to solve systems of
linear equations in time $O(\log N)$ - in an unconventional
sense. Namely, the algorithm of \cite{HHL} generates the quantum state
$\ket{x}=\sum_{i=1}^N x_i \ket{i}$ with the coefficients $x_i$ being
equal to the values of variables in the solution $x=(x_1, x_2, \ldots, x_N)$
of the system $Ax=b$.

The Harrow-Hassidim-Lloyd algorithm among the most interesting new
results in quantum algorithms, because systems of 
linear equations have many applications in all fields of science.
For example, this algorithm has been used to design quantum algorithms
for solving differential equations \cite{LO,Berry}.


Besides $N$, the running time of the algorithms for systems of linear
equations (both classical and quantum algorithms) depends on another parameter
$\kappa$, the condition number of matrix $A$. The condition number is defined as
the ratio between the largest and the smallest singular value of $A$: 
$\kappa = \max_{i, j} \frac{|\mu_i|}{|\mu_j|}$ where $\mu_i$ are the singular
values of $A$.

In the case of sparse classical matrices, the best classical algorithm
runs in time $O(\sqrt{\kappa} N)$ \cite{S94} while the HHL quantum algorithm runs
in time $O(\kappa^2 \log N)$, with an exponentially better dependence 
on $N$ but worse-than-classical dependence on $\kappa$.

In this paper, we present a better quantum algorithm, 
with the running time $O(\kappa \log^{3} \kappa \log N)$. To construct our algorithm,
we introduce a new tool, the {\em variable-time quantum amplitude amplification}
which allows to amplify the success probability of quantum
algorithms in which some branches of the computation stop earlier
than other branches. The conventional amplitude amplification \cite{BHMT}
would wait for all branches to stop - possibly resulting in a substantial
inefficiency. Our new algorithm amplifies the success probability in
multiple stages and takes advantage of the parts of computation which stop
earlier. We expect that this new method will be useful for building other
quantum algorithms.

The dependence of our quantum algorithm for solving systems of linear equations
on $\kappa$ is almost optimal. Harrow et al. \cite{HHL} show that, unless $BQP=PSPACE$, 
time of $\Omega(\kappa^{1-o(1)})$ is necessary for generating the state $\ket{x}$ that
describes the solution of the system.

\section{Overview of main results}

\subsection{Variable time amplitude amplification}

Informally, our result is as follows.
Consider a quantum algorithm $\A$ which may stop at one of several times $t_1, \ldots, t_m$.
(In the case of systems of linear equations, these times corresponding to $m$ runs of eigenvalue 
estimation with increasing precision and increasing number of steps.)
To indicate the outcome, $\A$ has an extra register $O$ with 3 possible values: 0, 1 and 2.
1 indicates the outcome that should be amplified. 0 indicates that the computation has stopped at 
this branch but did not the desired outcome 1. 2 indicates that the computation at this branch has 
not stopped yet. 

Let $p_{i}$ be the probability of the algorithm stopping at time $t_i$ (with either the outcome 0 or outcome 1).
The average stopping time of $\A$ (the $l_2$ average) is 
\[ T_{av} = \sqrt{\sum_i p_i t^2_i} .\]
$T_{max}$ denotes the maximum possible running time of the algorithm (which is equal to $t_m$).
Let 
\[ \alpha_{good} \ket{1}_O \ket{\psi_{good}} + \alpha_{bad} \ket{0}_O \ket{\psi_{bad}} \]
be the algorithm's output state after all branches of the computation have stopped.
Our goal is to obtain $\ket{\psi_{good}}$ with a high probability.
Let $p_{succ}=|\alpha_{good}|^2$ be the probability of obtaining this state via 
algorithm $\A$.

Our main result is
\begin{Theorem}
\label{thm:variable-amplify}
We can construct a quantum algorithm $\A'$ invoking $\A$ several times, for total time 
\[ O\left(T_{max} \sqrt{\log T_{max}} + \frac{T_{av}}{\sqrt{p_{succ}}} \log^{1.5} T_{max}  \right)
\]
that produces a state 
$\alpha \ket{1}\otimes \ket{\psi_{good}} +\beta\ket{0} \otimes \ket{\psi'}$
with probability $|\alpha|^2 \geq 1/2$ as the output\footnote{The first bit of the output state
indicates whether we have the desired state $\ket{\psi_{good}}$ or not. Since $|\alpha|^2 \geq 1/2$, we get 
$\ket{\psi_{good}}$  with probability at least 1/2.}
\end{Theorem}

In contrast, the usual amplitude amplification \cite{BHMT} would run for time $O(\frac{T_{max}}{\sqrt{p_{succ}}})$.
Our algorithm $\A'$ provides an improvement, whenever $T_{av}$ is substantially smaller than $T_{max}$.
By repeating $\A'$ $O(\log \frac{1}{\epsilon})$ times, we can obtain $\ket{\psi_{good}}$ with
a probability at least $1-\epsilon$.

Our algorithm $\A'$ is optimal, up to the factor of $\log T_{max}$. If the algorithm $\A$ has just one stopping time 
$T=T_{av}=T_{max}$, then amplitude amplification cannot be performed with fewer than $O(\frac{T}{\sqrt{p_{succ}}})$ steps.
Thus, the term of $\frac{T_{av}}{\sqrt{p_{succ}}}$ is necessary. The term $T_{max}$ is also necessary because, in some branch
of computation, $\A$ can run for $T_{max}$ steps.  

More details are given in section \ref{sec:variable}.
First, in subsection \ref{sec:model}, we give a precise definition of how a quantum algorithm
could stop at different times.
Then, in subsections \ref{sec:tools} and \ref{sec:var-alg}, we give a proof
of Theorem \ref{thm:variable-amplify}.

\subsection{Systems of linear equations}

We consider solving a system of linear equations $Ax=b$ where
$A=(a_{ij})_{i, j\in[N]}$, $x=(x_i)_{i\in [N]}$, $b=(b_i)_{i\in [N]}$.
We assume that $A$ is Hermitian. 
As shown in \cite{HHL}, this assumption is without the loss of generality.

Let $\ket{v_i}$ be the eigenvectors of $A$ and $\lambda_i$ be their eigenvalues.
Similarly to \cite{HHL}, we assume that all $\lambda_i$ satisfy
$\frac{1}{\kappa}\leq \lambda_i \leq 1$ for some known $\kappa$.
We can then transform the state $\ket{b}=\sum_{i=1}^n b_i \ket{i}$
into $\ket{x}=\sum_{i=1}^n x_i \ket{i}$ as follows:
\begin{enumerate}
\item
If, in terms of eigenvectors $\ket{v_i}$ of $A$, we have
$\ket{b}= \sum_i c_i \ket{v_i}$, then $\ket{x}= \sum_i \frac{c_i}{\lambda_i} \ket{v_i}$.
\item
By eigenvalue estimation, we can create the state $\ket{b'}=\sum_i c_i \ket{v_i} 
\ket{\tilde{\lambda}_i}$ where $\tilde{\lambda}_i$ are the estimates of the true 
eigenvalues. 
\item
We then create the state 
\begin{equation}
\label{eq:create} 
\ket{b''} = \sum_i c_i \ket{v_i} \ket{\tilde{\lambda}_i} 
\left( \frac{1}{\kappa\tilde{\lambda}_i}  \ket{1}+
\sqrt{1-\frac{1}{\kappa^2\tilde{\lambda}^2}} \ket{0} \right) . 
\end{equation}
Conditional on the last bit being 1, the rest of state is
$\sum_i \frac{c_i}{\tilde{\lambda}_i} \ket{v_i} \ket{\tilde{\lambda}_i}$
which can be turned into an approximation of $\ket{x}$ by running 
eigenvalue estimation in reverse and uncomputing $\tilde{\lambda}_i$.
\item
We then amplify the part of state which has the last qubit equal to 1 
(using amplitude amplification) and obtain a good approximation 
of $\ket{x}$ with a high probability.
\end{enumerate}

\begin{Theorem}
\cite{HHL}
Let $C$ be such that the evolution of the Hamiltonian $H$ for time $T$ can be simulated
in time $C\min(T, 1)$.
Then, we can generate $\ket{\psi'}$ satisfying $\|\psi-\psi'\| \leq \epsilon$
in time $(\frac{C \kappa^2}{\epsilon})$.
\end{Theorem}

The main term in the running time, $\kappa^2$ is generated as a product of two $\kappa$'s.
First, for $\|\psi-\psi'\| \leq \epsilon$, it suffice that the estimates $\tilde{\lambda}_i$ satisfy
$|\lambda_i-\tilde{\lambda}_i| = O(\epsilon \tilde{\lambda}_i)$. 
Since $\lambda_i=\Omega(1/\kappa)$, this means 
$|\lambda_i-\tilde{\lambda}_i| = O(\frac{\epsilon}{\kappa})$. 
To estimate $\lambda_i$ within error $O(\frac{\epsilon}{\kappa})$, we
need to run $H$ for time $O(\frac{\kappa}{\epsilon})$.
Second, for amplitude amplification, we may need to repeat 
the algorithm generating $\ket{b''}$ $O(\kappa)$ times - resulting in the total
running time $O(\kappa^2/\epsilon)$. 

For eigenvalue estimation, the worst case is when all of most of $\lambda_i$ are small
(of order $\Theta(1/\kappa)$). Then, $|\lambda_i-\tilde{\lambda}_i| = \Theta(\frac{\epsilon}{\kappa})$. 
and eigenvalue estimation with the right precision indeed requires time
$\Theta(\frac{\kappa}{\epsilon})$.

For amplitude amplification, the worst case is if most or all of $\lambda_i$ are
large (constant). Then, the coefficients $\frac{1}{\kappa\tilde{\lambda}_i}$ can be 
of order $\Theta(1/\kappa)$ and $\Theta(\kappa)$ repetitions are required for 
amplitude amplification.

We now observe that the two $\Theta(\kappa)$'s appear in the opposite cases. 
One of them appears when $\lambda_i$ is small 
($\lambda_i\approx \kappa$) but the other appears  
when $\lambda_i$ is large ($\lambda_i\approx 1$).

If all eigenvalues are of roughly similar magnitude (e.g., $\lambda\in [a, 2a]$ 
for some $a$), the running time becomes $O(\kappa/\epsilon)$ because we 
can do eigenvalue estimation in time to error $\epsilon a$ in $O(1/a \epsilon)$ 
and, for eigenvalue amplification, it suffices to repeat the generation
of $\ket{b''}$ $O(\kappa a)$ times (since the amplitude of $1$ in the last qubit of $\ket{b'}$ is
at least $\frac{1}{\kappa a}$ for every $v_i$).
Thus, the running time is
\[ O\left(\frac{1}{a \epsilon}\right) \cdot O(\kappa a) = O\left( \frac{\kappa}{\epsilon} \right) .\]
The problem is to achieve a similar running time
in the general case (when the eigenvalues $\lambda_i$ can range 
from $\kappa$ to 1).

To do that, we first design a version of eigenvalue estimation in which some branches of computation (corresponding to 
eigenvectors with larger eigenvalues $\lambda_i$) terminate earlier than others.
Namely, we start by running it for $O(1)$ steps. If we see that the estimate $\tilde{\lambda_i}$ 
for the eigenvalue is such that the allowed error $O(\epsilon \tilde{\lambda}_i)$ is more
than the expected error of the current run of eigenvalue estimation, we stop. 
Otherwise, we run eigenvalue estimation again, doubling its running time. 
This doubles the precision achieved by eigenvalue estimation. 
We continue this until the precision of current estimate becomes better than the 
allowed error of $O(\epsilon \tilde{\lambda}_i)$.

This gives a quantum algorithm in which different branches of computation stop at different times. 
We apply our variable-time amplitude amplification to this quantum algorithm. This gives us
\begin{Theorem}
\label{thm:linear-new}
Let $C$ be such that the evolution of the Hamiltonian $H$ for time $T$ can be simulated
in time $C\min(T, 1)$.
Then, we can generate $\ket{\psi'}$ satisfying $\|\psi-\psi'\| \leq \epsilon$
in time 
\[ O\left(\frac{C \kappa \log^3 \frac{\kappa}{\epsilon}}{\epsilon^3} \log^2 \frac{1}{\epsilon}\right). \] 
\end{Theorem}
We give more details in section \ref{sec:linear}.

\section{Variable-time amplitude amplification}
\label{sec:variable}

\subsection{Model}
\label{sec:model}

How can a quantum algorithm have different branches of computation stopping at different times?
We start by giving a precise definition of that.

We require the state space of $\A$ to be of the form 
$\H=\H_{o}\otimes \H_{c}$ be the Hilbert space of $\A$, consisting of the 0-1-2 valued 
outcome register $\H_o$ and the rest of the Hilbert space $\H_c$. 
Let $\ket{\psi_1}, \ldots, \ket{\psi_m}$ be the states of $\A$ at times $t_1, \ldots, t_m$.
We insist on the following consistency requirements. 
\begin{enumerate}
\item
For each $i\in\{1, \ldots, m\}$, the description of the algorithm must define 
a subspace $\H_i$ of $\H_o$ in which the computation has stopped.
Those subspaces must satisfy 
\[ \H_1\subseteq \H_2 \ldots \subseteq \H_m =\H_c .\]
\item
The state $\ket{\psi_i}$ can be expressed as
\[ \ket{\psi_i} = \alpha_{i, 0} \ket{0}\otimes \ket{\psi_{i,0}} + 
\alpha_{i, 1} \ket{1}\otimes \ket{\psi_{i,1}} + 
\alpha_{i, 2} \ket{2}\otimes \ket{\psi_{i,2}} ,\]
with $\ket{\psi_{i, 0}}\in \H_i$, $\ket{\psi_{i, 1}}\in \H_i$ and $\ket{\psi_{i, 2}}\in \H_o \cap (\H_i)^{\perp}$.
(When $i=m$, we have $\ket{\psi_{m, 0}}=\ket{\psi_{bad}}$, $\ket{\psi_{m, 1}}=\ket{\psi_{good}}$, $\ket{\psi_{m, 2}}=\overrightarrow{0}$.)
\item
We must have
\[ P_{H_i} \ket{\psi_{i+1, 0}} = \ket{\psi_{i, 0}} \mbox{~and~} 
P_{H_i} \ket{\psi_{i+1, 1}} = \ket{\psi_{i, 1}} .\]
That is, the part of the state where the computation stopped at time $t_i$ should not change after that.
\end{enumerate}
{\em The success probability} of $\A$ is $p_{succ} = |\alpha_{m, 1}|^2$.
We also define $p_{succ, i} = |\alpha_{i, 1}|^2$, the probability
of $\A$ succeeding before time $t_i$.
The probability of $\A$ stopping at time $t_i$ or earlier is
\[ p_{stop, \leq i} = | \alpha_{i, 0} |^2 + | \alpha_{i, 1} |^2 .\]
The probability of $\A$ stopping at exactly time $t_i$ is 
$p_{stop, 1}=p_{stop, \leq 1}$ for $i=1$ and $p_{stop, i}=p_{stop, \leq i} - 
p_{stop, \leq i-1}$ for $i>1$.
We will also use the probability of $\A$ stopping later than time $t_i$, 
defined as 
\[ p_{stop, >i} = |\alpha_{i, 2}|^2 = 1-p_{stop, \leq i}. \]
The average stopping time of $\A$ (the $l_2$ average) is 
\[ T_{av} = \sqrt{\sum_i p_i t^2_i} .\]
The maximum stopping time of $\A$ is $T_{max}=t_m$.
Our goal is to amplify the success probability to $\Omega(1)$, by running $\A$ for time 
$O\left(T_{max} \log^{0.5} T_{max}+ \frac{T_{av}}{\sqrt{p_{succ}}} \log^{1.5} T_{max}  \right)$.

\subsection{Tools}
\label{sec:tools}

Our variable-time amplitude amplification uses two subroutines.
Thr first is a result by Aaronson and Ambainis \cite{AA} 
who gave a tighter analysis of the usual amplitude amplification algorithm \cite{BHMT}.

We say that an algorithm $\A$ produces a quantum state $\ket{\psi}$ with probability $p$ if the following is true:
\begin{itemize}
\item
The algorithm has two output registers $R$ and $S$ (and, possibly some more auxiliary registers);
\item
Measuring $R$ gives 1 with probability $p$ and, conditional on this measurement result, the $S$ register is in state $\ket{\psi}$.
\end{itemize}

\begin{Lemma}
\label{lem:AA}
\cite{AA}
Let $\A$ be a quantum algorithm that outputs a state $\ket{\psi}$
with probability\footnote{\cite{AA} requires the probability to be exactly
$\epsilon$ but the proof works without changes if the probability is less than
the given $\epsilon$.} $\delta\leq \epsilon$ where $\epsilon$ is known.
Furthermore, let 
\begin{equation}
\label{eq:mconstraint} 
m\leq \frac{\pi}{4 \arcsin \sqrt{\epsilon}} - \frac{1}{2} .
\end{equation}
Then, there is an algorithm $\A'$ which uses $2m+1$ calls to $\A$ and $\A^{-1}$
and outputs a state $\ket{\psi}$ with probability 
\begin{equation}
\label{eq:aa} 
\delta_{new}\geq \left( 1-\frac{(2m+1)^2}{3} \delta \right) (2m+1)^2 \delta .
\end{equation}
\end{Lemma}  

The algorithm $\A'$ is just the standard amplitude amplification \cite{BHMT}
but its analysis is tighter.
According to the usual analysis, amplitude amplification 
increases the success probability from $\delta$ to $\Omega(1)$ 
in $2m+1=O(\frac{1}{\sqrt{\delta}})$ 
repetitions. In other words, $2m+1$ repetitions increase the success
probability $\Omega((2m+1)^2)$ times. Lemma \ref{lem:AA} achieves an
increase of almost $(2m+1)^2$ times, without the big-$\Omega$ factor.

This is useful if we have an algorithm with $k$ levels of amplitude
amplification nested one inside another. 
Then, with the usual amplitude amplification,
a big-$\Omega$ constant of $c$ would result in a $c^{k}$ 
factor in the running time. Using Lemma \ref{lem:AA} avoids that.

Our second subroutine is a version of amplitude estimation from \cite{A08}.

\begin{Theorem}
\cite{BHMT,A08}
\label{thm:est}
There is a procedure {\bf Estimate}$(\A, c, p, k)$ which, given 
a constant $c$, $0<c\leq 1$ and a quantum algorithm $\A$
(with the promise that the probability $\epsilon$ that 
the algorithm $\A$ outputs 1 is either 0 or at least a 
given value $p$)
outputs an estimate $\tilde{\epsilon}$ of
the probability $\epsilon$ such that, 
with probability at least $1-\frac{1}{2^k}$, 
we have
\begin{enumerate}
\item[(i)]
$|\epsilon-\tilde{\epsilon}|<c \tilde{\epsilon}$ if $\epsilon\geq p$;
\item[(ii)] 
$\tilde{\epsilon}=0$ if $\epsilon=0$.
\end{enumerate}
The procedure {\bf Estimate}$(\A, c, p, k)$ uses the expected number of
\[ \Theta\left(k\left(1+\log \log \frac{1}{p}\right) 
\sqrt{\frac{1}{\max(\epsilon, p)}} \right) \]
evaluations of $\A$.
\end{Theorem}

\subsection{The state generation algorithm}
\label{sec:var-alg}

We now describe our state generation algorithm. Without the loss of generality,
we assume that the stopping times of $\A$ are $t_i=2^i$ for $i\in\{0, \ldots, m\}$ for some $m$.
We present a sequence of algorithms $\A_i$, with the algorithm $\A_i$ generating an 
approximation of the state 
\[ \ket{\psi'_{i}} = \frac{\alpha_{i, 1}}{\sqrt{|\alpha_{i, 1}|^2+|\alpha_{i, 2}|^2}} 
\ket{1}\otimes \ket{\psi_{i, 1}} + \frac{\alpha_{i, 2}}{\sqrt{|\alpha_{i, 1}|^2+|\alpha_{i, 2}|^2}} 
\ket{2}\otimes \ket{\psi_{i, 2}} ,\] 
in the following sense: the algorithm $\A_i$ outputs a state 
\begin{equation}
\label{eq:ri} 
\ket{\psi''_i} = \sqrt{r_i} \ket{\psi'_i} + \sqrt{1-r_i} \ket{0} \otimes \ket{\phi_i} 
\end{equation}
for some $\ket{\phi_i}$ and some $r_i$ satisfying $r_i \geq 1/9m$.
(To avoid the problem with nested amplitude amplification described in section
\ref{sec:tools}, we only require $r_i \geq 1/9m$ instead of $r_i = \Omega(1)$.)

The algorithm $\A_i$ uses $\A_{i-1}$ as the subroutine. It is defined in two steps.
First, we define an auxiliary algorithm $\B_i$. 

\begin{Algorithm}
\begin{enumerate}
\item
If $i=0$, $B_i$ runs $\A$ for 1 step and outputs the output state of $\A$.
\item
If $i>0$, $B_i$ runs $\A_{i-1}$ which outputs $\ket{\psi''_{i-1}}$. $\B_i$ then executes
$\A$ for time steps from $2^{i-1}$ to $2^i$ on the parts of the state $\ket{\psi''_{i-1}}$ where
the outcome register is 2 (the computation is not finished).
\end{enumerate}
\caption{Algorithm $\B_i$}
\label{alg:bi}
\end{Algorithm}

Let $p_i={\bf Estimate} (\B_i, c, \frac{1}{\kappa}, \log m + 5)$. 
Then, $\A_i$ is as follows.

\begin{Algorithm}
\begin{enumerate}
\item
If $p> \frac{1}{9m}$, $\A_i=\B_i$.
\item
If $p\leq \frac{1}{9m}$, $\A_i={\bf Amplify} (B_i, k)$ for the smallest $k$ 
satisfying $\frac{1}{9m} \leq (2k+1)^2 p \leq \frac{1}{m}$.
\end{enumerate}
\caption{Algorithm $\A_i$}
\label{alg:ai}
\end{Algorithm}

The overall algorithm $\A'$ is given as Algorithm \ref{alg:aprime}.
\begin{Algorithm}
\begin{enumerate}
\item
Run {\bf Estimate} to obtain $p_0={\bf Estimate} (\B_0, c, \frac{1}{\kappa}, \log m + 5)$. 
\item
For each $i=1, 2, \ldots, m$:
\begin{enumerate}
\item
Use $p_{i-1}$ to define $\A_i$ and $\B_i$.
\item
If $i<m$, run {\bf Estimate} to obtain 
$p_i={\bf Estimate} (\B_i, c, \frac{1}{\kappa}, \log m + 5)$.  
\end{enumerate}
Amplify $\A_m$ to the success probability at least 1/2 and output the output state
of the amplified $\A_m$.
\end{enumerate}
\caption{Algorithm $\A'$}
\label{alg:aprime}
\end{Algorithm}

We now analyze the running times of algorithms $\A_i$. 
Let $T_i$ denote the running time of $\A_i$.
Let $r_i$ be as defined in equation (\ref{eq:ri}) and let $r'_i$
be a similar quantity for the output state of $\B_i$.
Then, we have
\begin{Lemma}
\label{lem:lem1}
\begin{equation}
\label{eq:rec} 
T_i \leq \left( 1+\frac{1}{3m-1} \right) \frac{\sqrt{r_i}}{\sqrt{r'_i}} \left( T_{i-1}
+2^{i-1} \right) .
\end{equation}
\end{Lemma}

\proof
The running time of $\B_i$ is $T_{i-1}+2^{i-1}$. If $\A_i=\B_i$, then 
the running time of $\A_i$ is the same and, also $r_i=r'_i$ (because the
two algorithms output the same state). If $\A_i$ is an amplified version of $\B_i$,
then:
\begin{enumerate}
\item
The running time of $\A_i$ is $(2k+1)(T_{i-1}+2^{i-1})$.
\item
By Lemma \ref{lem:AA}, we have $r_i \geq (1-\frac{1}{3m}) (2k+1)^2 r'_i$ which implies
\[ (2k+1) \leq (1+\frac{1}{3m-1}) \frac{\sqrt{r_i}}{\sqrt{r'_i}} .\]
\end{enumerate} 
\qed

Applying (\ref{eq:rec}) recursively, we get
\begin{equation}
\label{eq:tm} 
T_m \leq \left( 1+\frac{1}{3m-1} \right)^m \sum_{i=1}^m \left( \prod_{j=i}^m
\frac{\sqrt{r_i}}{\sqrt{r'_i}}\right) 2^{i-1} .
\end{equation}
The first multiplier, $\left( 1+\frac{1}{3m-1} \right)^m$ can be upper-bounded
by a constant. We now bound the product $\prod_{j=i}^m \frac{\sqrt{r_i}}{\sqrt{r'_i}}$.

\begin{Lemma}
\label{lem:ratio}
\[ \prod_{j=i}^m \frac{\sqrt{r_i}}{\sqrt{r'_i}} \leq 3 \left( 
1+ \sqrt{\frac{p_{stop, >i}}{p_{succ}} } \right) .\]
\end{Lemma}
  
\proof
We consider the quantities 
\[ o_j = |\lbra 1 \otimes \psi_{i, 1} | \psi''_i \rket|^2 \]
for $j=i, i+1, \ldots, m$. For $j=i$, we have 
\begin{equation}
\label{eq:overlap} 
o_i = r_i |\lbra 1 \otimes \psi_{i, 1} | \psi'_i \rket|^2 =
r_i \frac{|\alpha_{i, 1}|^2}{|\alpha_{i, 1}|^2+|\alpha_{i, 2}|^2} =
r_i \frac{p_{succ, i}}{p_{succ, i}+p_{stop, >i}} .
\end{equation}
For $j>i$, we have $o_j = o_{j-1} \frac{r_i}{r'_i}$ because amplification
increases the probability of the "good" part of the state 
(which includes $\ket{1 \otimes \psi_{i, 1}}$) $\frac{r_i}{r'_i}$ times.
Finally, we have
\[ o_m = r_m \frac{p_{succ, i}}{p_{succ}} \]
which follows similarly to (\ref{eq:overlap}). Putting all of this together,
we have  
\[ \prod_{j=i}^m \frac{r_i}{r'_i} = \frac{o_m}{o_i} =
\frac{r_m}{r_i} \cdot \frac{p_{succ, i}+p_{stop, >i}}{p_{succ, i}} .\]
By taking the square roots from both sides and observing that  
$\frac{r_m}{r_i}$ is at most 9 (because 
$r_m \leq \frac{1}{m}$ and $r_i \geq \frac{1}{9m}$), we get
\[ \prod_{j=i}^m \frac{\sqrt{r_i}}{\sqrt{r'_i}} \leq 3 \sqrt{
1+\frac{p_{stop, >i}}{p_{succ}} } .\]
The Lemma follows by using $\sqrt{1+x}\leq 1+\sqrt{x}$.
\qed

By applying Lemma \ref{lem:ratio} to each term in (\ref{eq:tm}), we get
\[ T_m \leq C \sum_{i=1}^m \left( 1+ \sqrt{\frac{p_{stop, >i}}{p_{succ}} } \right) 2^{i-1}
= C \sum_{i=1}^m 2^{i-1} + C \frac{\sum_{i=1}^m 2^{i-1} \sqrt{p_{stop, >i}}}{\sqrt{p_{succ}}}
.\]
The first sum can be upper bounded by $2^i=O(T_{max})$.
For the second sum, in its numerator, we have
\[ \sum_{i=1}^m 2^{i-1} \sqrt{p_{stop, >i}} = \sum_{i=1}^m \sqrt{2^{2i-2} p_{stop, >i}}
\leq m T_{av} = T_{av} \log T_{max} \]
where the inequality follows because each term $\sqrt{2^{2i-2} p_{stop, >i}}$ is at most $T_{av}$.
Thus, the algorithm $\A_m$ runs in time 
\[ O \left( T_{max} + \frac{T_{av}}{\sqrt{p_{succ}}} \log T_{max} \right) .\]
The algorithm $\A'$ amplifies $\A_m$ from a success probability of 
$r_m\geq \frac{1}{9m}$ to a success probability $\Omega(1)$.
This increases the running time by a factor of $O(\sqrt{m}) = O(\sqrt{\log T_{max}})$.

\section{Faster algorithm for solving systems of linear equations}
\label{sec:linear}

\subsection{Unique-answer eigenvalue estimation}

For our algorithm, we need a version of eigenvalue estimation 
that is guaranteed to output exactly the same estimate with a high probability. 
\comment{This is due to the fact that, in our algorithm (which we describe in
section \ref{sec:main}), we will need to make a choice
(between stopping the algorithm and continuing it), depending on
the eigenvalue estimate and, after the choice, we will need to
uncompute the eigenvalue estimate.  
}
The standard version of eigenvalue estimation \cite[p. 118]{KLM} runs
$U=e^{-iH}$ up to $2^n$ times and, if the input is an eigenstate
$\ket{\psi}: H\ket{\psi}=\lambda\ket{\psi}$, outputs 
$x\in\{0, \frac{\pi}{2^n}, \frac{2\pi}{2^n}, \ldots, \frac{(2^n-1) \pi}{2^n}\}$ 
with probability
\begin{equation}
\label{eq:est} 
p(x)=\frac{1}{2^{2n}} \frac{\sin^2 2^n (\lambda - x)}{
\sin^2 (\lambda - x)} 
\end{equation}
(equation (7.1.30) from \cite{KLM}). 
We now consider an algorithm that runs the standard eigenvalue estimation
$k_{uniq}$ times and takes the most frequent answer $x_{maj}$.

\begin{Lemma}
\label{lem:unique}
For $k_{uniq} = O(\frac{1}{\epsilon^2} \log \frac{1}{\epsilon})$, we have
\begin{enumerate}
\item
If $|\lambda-x| \leq \frac{1 - \epsilon}{2^{n+1}}$, then
$Pr[x_{maj}=x] \geq 1-\epsilon$.
\item
If $\lambda\in [x+\frac{1 - \epsilon}{2^{n+1}} , x+\frac{1+ \epsilon}{2^{n+1}} ]$, then
$Pr[x_{maj}\in \{x, x+1\}] \geq 1-\epsilon$.
\end{enumerate}
\end{Lemma} 

\proof
In the first case, (\ref{eq:est}) is at least $(1+\epsilon)\frac{4}{\pi^2}$
for the correct $x$ and less than $\frac{4}{\pi^2}$ for any other $x$.
Repeating eigenvalue estimation $O(\frac{1}{\epsilon^2})$ times and taking the majority 
allows to distinguish the correct $x$ with a fixed probability (say 3/4) and
repeating it $O(\frac{1}{\epsilon^2} \log \frac{1}{\epsilon})$ times allows to determine the 
correct $x$ with a probability at least $1-\epsilon$.

In the second case, the two values $x$ and $x+1$ are output with probability at least
$(1-\epsilon) \frac{4}{\pi^2}$ each. In contrast, for any other 
$y=\frac{m \pi}{2^n}$, $m\in\{0, 1, \ldots, 2^n-1\}$, 
we have
\[ |y- \lambda|\geq \frac{1-\epsilon}{2^{n+1}} \pi + \frac{1}{2^n} \pi =
\frac{3-\epsilon}{2^{n+1}} \pi .\]
This implies
\[ p(y) \leq \frac{1}{2^{2n}} \frac{1}{\sin^2 \frac{(3-\epsilon)}{2^{n+1}} \pi } =
(1+o(1)) \frac{4}{(3-\epsilon)^2 \pi^2} .\]
Thus, there is a constant gap between $p(x)$ or $p(x+1)$ and $p(y)$ for any other $y$.
In this case, taking majority of $O(\log \frac{1}{\epsilon})$ runs of
eigenvalue estimation is sufficient to produce $x$ or $x+1$ with a probability
at least $1-\epsilon$.  
\qed

We refer to this algorithm as {\bf UniqueEst}$(H, 2^n, \epsilon)$.

When we use {\bf UniqueEst} as a subroutine in
algorithm \ref{alg:main}, we need the answer to be unique (as in the first case) and 
not one of two high-probability answers (as in the second case).
To deal with that, we will replace $H$ with $H+\frac{\delta\pi}{2^n}I$ for a randomly chosen 
$\delta\in[0, 1]$. The eigenvalue becomes 
$\lambda'=\lambda+\frac{\delta\pi}{2^n}$ and, with probability $1-\epsilon$,
\[ \lambda'\in\left[\frac{x - \frac{1 - \epsilon}{2}}{2^n} \pi, 
\frac{x + \frac{1 - \epsilon}{2}}{2^n} \pi \right] \]
for some integer $x$. This allows to achieve the first case for all eigenvalues, except a small
random fraction of them.

\comment{\subsection{Variable time amplitude estimation}

We now define amplitude amplification for quantum algorithms in which different branches terminate at
different times.}

\subsection{Main algorithm}
\label{sec:main}

We now show that Theorem \ref{thm:variable-amplify} implies our main result, Theorem \ref{thm:linear-new}.
We start by describing a variable running time Algorithm \ref{alg:amplify}.
This algorithm uses the following registers: 
\begin{itemize}
\item
The input register $I$ which holds the input state $\ket{x}$ (and is also used
for the output state);
\item
The outcome register $O$, 
with basis states $\ket{0}, \ket{1}$ and $\ket{2}$ (as described in the setup for
variable-time amplitude amplification);
\item
The step register $S$, with basis
states $\ket{1}$, $\ket{2}$, $\ldots$, $\ket{2m}$ (to prevent interference between
various branches of computation).
\item
The estimation register $E$, which is used for eigenvalue estimation
(which is a subroutine for our algorithm).
\end{itemize}
$\H_I$, $\H_O$, $\H_S$ and $\H_E$ denote the Hilbert spaces of the
respective registers.

From now on, we refer to $\epsilon$ appearing in Theorem \ref{thm:linear-new} as $\epsilon_{final}$.
$\epsilon$ without a subscript is an error parameter for subroutines of algorithm \ref{alg:amplify}
(which we will choose at the end of the proof so that the overall error in the output state
is at most $\epsilon_{final}$).

\begin{Algorithm}
{\bf Input:} parameters $x_1, \ldots, x_m\in[0, 1]$, Hamiltonian $H$.
\begin{enumerate}
\item
Initialize $O$ to $\ket{2}$, $S$ to $\ket{1}$ and $E$ to $\ket{0}$. Set $j=1$.
\item
Let $m=\lceil \log_2 \frac{\kappa}{\epsilon}\rceil$.
\item
Repeat until $j>m$:

{\bf Stage $j$:}
\begin{enumerate}
\item
Let $H'=H+\frac{x_j \pi}{2^j} I$.
Using the registers $I$ and $S$, run {bf UniqueEst}$(H', 2^j, \epsilon)$.
Let $\lambda'$ be the estimate output by {\bf UniqueEst} and let $\lambda=\lambda'-\frac{x_j\pi}{2^j}$.
\item
\label{st:term}
If $\epsilon \lambda >  \frac{1}{2^{j+1}}$,
perform the transformation
\begin{equation}
\label{eq:transform} \ket{2}_O \otimes \ket{1}_S
\rightarrow \frac{1}{\kappa \lambda}  \ket{1}_O\otimes \ket{2j}_S+
\sqrt{1-\frac{1}{(\kappa\lambda)^2}} \ket{0}_O\otimes \ket{2j}_S .
\end{equation}
\item
Run {\bf UniqueEst} in reverse, to erase the intermediate information.
\item
\label{st:check}
Check if the register $E$ is in the correct initial state $\ket{0}_E$. If not, apply  
$\ket{2}_O\otimes \ket{1}_S \rightarrow \ket{0}_O\otimes \ket{2j+1}_S$
on the outcome register $O$.
\item
If the outcome register $O$ is in the state $\ket{2}$, 
increase $j$ by 1 and go to step 2.
\end{enumerate}
\end{enumerate}
\caption{State generation algorithm}
\label{alg:amplify}
\end{Algorithm}

Our main algorithm is Algorithm \ref{alg:main} which consists of applying 
variable-time amplitude amplification to Algorithm \ref{alg:amplify}.

\begin{Algorithm}
{\bf Input:} Hamiltonian $H$.
\begin{enumerate}
\item
Generate uniformly random $x_1, \ldots, x_m\in[0, 1]$.
\item
Apply variable-time amplitude amplification to Algorithm \ref{alg:amplify}, with $H$ and $x_1, \ldots, x_m$ as the input.
\item
Apply a transformation mapping $\ket{2j}_S \rightarrow \ket{j}_S$ to the $S$ register. 
After that, apply Fourier transform $F_m$ to the $S$ register and measure. 
If the result is 0, output the state in the $I$ register.
Otherwise, stop without outputting a quantum state.
\end{enumerate}
\caption{Main algorithm}
\label{alg:main}
\end{Algorithm}

We claim that, conditional on the output register being $\ket{1}_O$,
the output state of Algorithm \ref{alg:amplify} is close to
\begin{equation}
\label{eq:ideal} 
\ket{\psi_{ideal}}= \sum_i \alpha_i \ket{v_i}_I \otimes
\left( \frac{1}{\kappa \lambda_i}  \ket{1}_O\otimes \ket{2j_i}_S \right) .
\end{equation}
Variable-time amplitude amplification then generates a state that is
close to $\frac{\ket{\psi_{ideal}}}{\|\psi_{ideal}\|}$. 
Fourier transform in the last step of algorithm \ref{alg:main}
then effectively erases the $S$ register. Conditional on $S$ being in $\ket{0}_S$ after
the Fourier transform, the algorithm's output state is close to 
our desired output state $\frac{\ket{x}}{\|x\|}$, where
\[ \ket{x}= \sum_i \alpha_i \ket{v_i}_I .\]
Finally, performing Fourier transform and measuring
produces $\ket{0}_S$ with probability $1/m$. 
Because of that, the success probability of algorithm \ref{alg:main}
needs to be amplified. This adds a factor of $O(\sqrt{m})$ to the running time,
if we would like to obtain the result state with probability $\Omega(1)$ and
a factor of $O(\sqrt{m} \log \frac{1}{\epsilon})$ if we would like to obtain
it with probability at least $1-\epsilon$.

{\bf Approximation guarantees.}
We now give a formal proof that the output state of Algorithm \ref{alg:amplify} is close
to the desired output state (\ref{eq:ideal}).

Let $\ket{v_i}$ be an eigenvector and $\lambda_i$ be an eigenvalue.
For each $j$, the unique-value eigenvalue estimation either outputs 
one estimate $\tilde{\lambda}_{i, j}$ or one of two estimates
$\tilde{\lambda}_{i, j}$ and $\tilde{\lambda}_{i, j}-\frac{1}{2^j}$
with a high probability (at least $1-\epsilon$). 
Let $j_i$ be the smallest $j$ for which the estimate 
$\tilde{\lambda}=\tilde{\lambda}_{i, j}$
satisfies the condition $\epsilon \tilde{\lambda} \geq \frac{1}{2^{j+1}}$
in step \ref{st:term}.
We call $v_i$ and $\lambda_i$ {\em good} if, for $j=j_i$ the unique-value eigenvalue estimation
outputs one estimate $\tilde{\lambda}_{i, j}$ with a high probability.
Otherwise, we call $\lambda_i$ {\em bad}. For both good and bad $\lambda_i$,
we denote $\tilde{\lambda}_i=\tilde{\lambda}_{i, j_i}$.

We claim that the part of final state Algorithm \ref{alg:amplify} that has $\ket{1}$ in the output register $O$ is close to 
\[ \ket{\psi'}= \sum_i \alpha_i \ket{v_i}_I \otimes
\left( \frac{1}{\kappa \tilde{\lambda}_i}  \ket{1}_O\otimes \ket{2j_i}_S \right)
\]
and $\ket{\psi'}$ is, in turn, close to the state $\ket{\psi_{ideal}}$
defined by equation (\ref{eq:ideal}).

The next two lemmas quantify these claims. Let 
\[ \delta = \sum_{i: \lambda_i \mbox{~bad}} |\alpha_i|^2 
\]
quantify the size of the part of the state $\ket{\psi'}$ that consists of 
bad eigenvectors.

\begin{Lemma}
\label{lem:close}
Let $\ket{\psi}$ be the output state of Algorithm \ref{alg:amplify} and
let $P_1$ be the projection to the subspace where the outcome register 
$O$ is in the state $\ket{1}$.
Then,  we have
\[ \|P_1\ket{\psi}-\ket{\psi'}\|\leq ((2m+37)\epsilon+30\delta) \|\psi'\| .\]
\end{Lemma}

\proof
In section \ref{sec:proofs}.
\qed

\begin{Lemma}
\label{lem:close1}
\[ \|\ket{\psi'}-\ket{\psi_{ideal}}\|\leq \frac{2 \epsilon}{1+2\epsilon} \|\psi_{ideal}\| .\]
\end{Lemma}

\proof
In section \ref{sec:proofs}.
\qed

When $x_1, \ldots, x_m\in[0, 1]$ are chosen uniformly at random, the probability of any 
given $v_i$ being bad is of order $O(\epsilon)$. Thus, $E[\delta]=O(\epsilon)$ and
\[ E\|P_1\ket{\psi}-\ket{\psi_{ideal}} \| = O(m \epsilon \|\psi_{ideal} \| ) \]
with the expectation taken over the random choice of $x_1, \ldots, x_m\in[0, 1]$.

To achieve an error of at most $\epsilon_{final}$, we choose
$\epsilon=\Theta(\epsilon_{final}/m)$. 

{\bf Running time.}
We now bound the running time of Algorithm \ref{alg:amplify}. We start with two lemmas
bounding the average running time $T_{av}$ and success probability $p_{av}$.

\begin{Lemma}
\label{lem:l2}
$T_{av}$, the $l_2$-average running time of Algorithm \ref{alg:amplify}, is of the order
\begin{equation}
\label{eq:lem-l2}
 O\left(\sqrt{ \sum_{i} |\alpha_i|^2 2^{2 j_i} k_{uniq}^2}\right) .
\end{equation}
 where $k_{uniq}$ is the quantity from Lemma \ref{lem:unique}.
\end{Lemma}

\proof
In section \ref{sec:proofs1}.
\qed

\begin{Lemma}
\label{lem:prob}
$p_{succ}$, the success probability of Algorithm \ref{alg:amplify}, is 
\begin{equation}
\label{eq:lem-prob} 
\Omega\left( \sum_i |\alpha_i|^2 \frac{\epsilon^2 2^{2j_i}}{\kappa^2} 
\right) .
\end{equation}
\end{Lemma}

\proof
In section \ref{sec:proofs1}.
\qed

By dividing the two expressions above one by another, we get

\begin{Corollary}
\label{cor:runtime}
\[ \frac{T_{av}}{\sqrt{p_{succ}}} = O\left( \frac{\kappa}{\epsilon} k_{uniq}\right) .\]
\end{Corollary}


By Theorem \ref{thm:variable-amplify}, the running time of algorithm \ref{alg:main} is
\[ O\left(T_{max} \sqrt{\log T_{max}} +\frac{T_{av}}{\sqrt{p_{succ}}} \log^{1.5} T_{max}\right). \]
Since $T_{max}=O(2^m)=O(\frac{\kappa}{\epsilon})$, we have 
$T_{max}\leq \frac{T_{av}}{\sqrt{p_{succ}}}$ and the running time is
\[ O\left(\frac{T_{av}}{\sqrt{p_{succ}}} \log^{1.5} T_{max} \right) = 
O\left( \frac{\kappa}{\epsilon} k_{uniq} \log^{1.5} \frac{\kappa}{\epsilon} \right) =
O\left( \frac{m \kappa}{\epsilon_{final}} k_{uniq} \log^{1.5} \frac{\kappa}{\epsilon}  \right),\]
with the 2nd equality following from $\epsilon=\Theta(\epsilon_{final}/m)$.
Since algorithm \ref{alg:main} needs to be repeated 
$O(\sqrt{m} \log \frac{1}{\epsilon_{final}})$
times, the overall running time is 
\[ 
O\left( \frac{m^{1.5} \kappa}{\epsilon_{final}}k_{uniq} \log^{1.5} \frac{\kappa}{\epsilon}  \log \frac{1}{\epsilon_{final}} \right) =
O\left( \frac{\kappa \log^{3} \frac{\kappa}{\epsilon} }{\epsilon^3_{final}} \log^2 \frac{1}{\epsilon_{final}} \right),\]
with the equality following from $m = O(\log \frac{\kappa}{\epsilon})$.

\subsection{Proofs of Lemmas about the quality of output state}
\label{sec:proofs}

\proof [of Lemma \ref{lem:close}]
Let $\ket{v_i}$ be an eigenstate of $\A$. Then, the eigenvalue
estimation leaves $\ket{v_i}$ unchanged (and produces an estimate
for the eigenvalue $\lambda_i$ in the $E$ register).
This means that the algorithm above maps $\ket{x}=\sum_i \alpha_i \ket{v_i}$ to
\[ \sum_i \alpha_i \ket{v_i}_I \otimes \ket{\phi_i}_{O, S, E} \]
where 
\[ \ket{\phi_i}_{O, S, E} = 
\ket{1}_O \otimes \ket{\phi'_i}_{S, E} + \ket{0}_O \otimes \ket{\phi''_i}_{S, E} .\]
We will show:
\begin{itemize}
\item
If $\ket{v_i}$ is good, then $\ket{\phi'_i}_{S, E}$ is close to 
$\frac{1}{\kappa \tilde{\lambda}_i} \ket{2j_i}_S \otimes \ket{0}_E$.
\item
If $\ket{v_i}$ is bad, then $\|\phi'_i\|$ does not become too large
(and, therefore, does not make too big contribution to 
$\|P_1 \ket{\psi}-\ket{\psi''}\|$).
\end{itemize}
These two statements are quantified by two claims below: Claim \ref{cl:good} and Claim \ref{cl:bad}.
The Lemma follows by combining these two claims and the fact that the sum
of $|\alpha_i|^2$ over all bad $i$ is equal to $\delta$.

Before proving Claims \ref{cl:good} and \ref{cl:bad}, we prove a claim that bounds $\tilde{\lambda_i}$
(and will be used in the proofs of both Claim \ref{cl:good} and Claim \ref{cl:bad}).

\begin{Claim}
\label{cl:doubling}
Let $j=j_i$. Then
\[ \frac{1}{\epsilon 2^{j+1}} \leq \tilde{\lambda}_i 
\leq \left( \frac{1}{\epsilon}+\frac{3}{2} \right) \frac{1}{2^j} .\]
\end{Claim}

\proof
The first inequality follows immediately. For the second inequality, since $j>j_i-1$, we
have 
\[ \tilde{\lambda}_{i, j-1} \leq \frac{1}{\epsilon 2^j} .\]
This means that the actual eigenvalue $\lambda$ satisfies 
\[ \lambda \leq (1+\epsilon) \frac{1}{\epsilon 2^j} = \frac{1}{\epsilon 2^j} + \frac{1}{2^j} \]
and 
\[ \tilde{\lambda}_{i, j} \leq  (1+\epsilon) \lambda \leq 
\frac{1}{\epsilon 2^j} + \frac{1}{2^j} + \frac{1}{2^{j+1}} . \]
\qed

As a consequence to this claim, we have
\[ \frac{1}{\tilde{\lambda}_i} \geq \left( \frac{2}{2+3\epsilon} \right) \epsilon 2^j .\]

\begin{Claim}
\label{cl:good}
If $\ket{v_i}$ is good, 
\[ \left\| \ket{\phi'_i} - \frac{1}{\kappa \tilde{\lambda}_i}  
\ket{1}_O \otimes \ket{2j_i}_S \otimes \ket{0}_E 
\right\|^2 \leq (2m+37) \epsilon C\]
where $C=(\frac{1}{\kappa\tilde{\lambda}_i})^2$. 
\end{Claim}

\proof
We express $\ket{\phi'_i} = \sum_{j} 
\ket{2j}_S \otimes \ket{\phi_{i, j}}_E$.
Furthermore, we group the terms of $\ket{\phi'_i}$ in a following way:
\[ \ket{\phi'_i} = \ket{\phi_{<}}+\ket{\phi_{=}}+\ket{\phi_{>}} \] 
where 
\[ \ket{\phi_{<}}=\sum_{j<j_i} \ket{2j}_S \otimes \ket{\phi_{i, j}}_E, \]
\[ \ket{\phi_{=}}=\otimes \ket{2j_i}_S \otimes \ket{\phi_{i, j_i}}_E, \]
\[ \ket{\phi_{>}}=\sum_{j>j_i}  \ket{2j}_S \otimes \ket{\phi_{i, j}}_E.\]
We have
\[ \left\| \ket{\phi'_i} - \frac{1}{\kappa \tilde{\lambda}_i}   \ket{2j_i}_S
\otimes \ket{0}_E \right\|^2 = \]
\[ \|\phi_{<}\|^2 + \left\| \ket{\phi_{=}} - 
\frac{1}{\kappa \tilde{\lambda}_i}  \ket{2j_i}_S
\otimes \ket{0}_E\right\|^2 + \|\phi_{>}\|^2 .\]
We first show that $\|\phi_{<}\|$ and $\|\phi_{>}\|$ are
not too large.

For $j<j_i$, the eigenvalue estimation
outputs an answer that is more than $\tilde{\lambda}_{i, j}$ 
with probability at most $\epsilon$. Therefore, the probability of step (\ref{st:term})
being executed is at most $\epsilon$.
Moreover, if this step is executed, 
the estimate $\lambda'$ for the eigenvalue is at least $\frac{1}{\epsilon 2^j}$.
Therefore, the coefficient of $\ket{1}_O$ in (\ref{eq:transform})
is 
\[ \frac{1}{\kappa \lambda'} \leq \frac{2^{j+1} \epsilon}{\kappa} .\]
By summing over all $j<j_i$, we get
\[ \|\phi_{<}\|^2 = \sum_{j<j_i} \|\phi'_{i, j}\|^2 
= \sum_{j<j_i} \left( \frac{2^{j+1} \epsilon}{\kappa} \right)^2 \epsilon \leq  
\frac{1}{3} \left( \frac{2^{j_i+1} \epsilon}{\kappa} \right)^2 \epsilon ,\]
with the inequality following from the formula for the sum of a geometric progression.
By using the right hand side of Claim \ref{cl:doubling}, we get
\[ \|\phi_{<}\|^2 \leq \frac{4\epsilon}{3} \left( 1+ \frac{3\epsilon}{2} \right)^2 C\]
where $C=(\frac{1}{\kappa\tilde{\lambda}_i})^2$.
If $\epsilon<0.1$, we can upper-bound this by $1.6 \epsilon C$.

For $j>j_i$, we have $\|\phi_{i, j}\|^2\leq \epsilon^{j-j_i}$.
(We only reach stage $j$ if, in every previous stage $k$, 
eigenvalue estimation outputs an estimate that is smaller than
$\tilde{\lambda}_{i}$. For each $k\in\{j_i, j_i+1, \ldots, j-1\}$,
this happens with probability at most $\epsilon$.)

Therefore,
\[ \|\phi_{>}\|^2 = \sum_{j>j_i} \|\phi'_{i, j}\|^2 \leq
\sum_{j>j_i} \left( \frac{2^{j+1} \epsilon}{\kappa} \right)^2 \epsilon^{j-j_i} \leq \]
\[ 4 \left( 1+ \frac{3\epsilon}{2} \right)^2 C
\sum_{j=1}^{\infty} (4\epsilon)^j =
16 \left( 1+ \frac{3\epsilon}{2} \right)^2 \frac{\epsilon}{1-4\epsilon} C\]
where the 2nd inequality follows from the right hand side of Claim \ref{cl:doubling}
and the last equality follows from the formula for the sum of a geometric progression.
If $\epsilon<0.1$, we can upper bound this by $36 \epsilon C$.
Thus, both $\|\phi_{<}\|^2$ and $\|\phi_{>}\|^2$ are small enough.
 
For $\ket{\phi_=}$, we first estimate the probability that algorithm reaches 
stage $j_i$.

\begin{Claim}
\label{cl:c1}
Algorithm \ref{alg:amplify} reaches stage $j_i$ with probability at least 
$1-2(m-1)\epsilon$.
\end{Claim}

\proof
For each $j<j_i$, the eigenvalue estimation may produce an incorrect answer  
with probability at most $\epsilon$.
This may lead to transformation (\ref{eq:transform}) being executed with probability at most $\epsilon$.
Moreover, this causes some disturbance for the next step,
when eigenvalue estimation is uncomputed. Let $\ket{\psi}$ be
the output of the eigenvalue estimation. We can split 
$\ket{\psi}=\ket{\psi'}+\ket{\psi''}$ where $\ket{\psi'}$ consists
of estimates $\lambda$ which are smaller than the one in the condition of step
\ref{st:term} and $\ket{\psi''}$ consists of estimates that are greater than or
equal to the one in the condition. Then, $\|\psi''\|^2\leq \epsilon$ and, conditional 
on outcome register being $\ket{2}$, the estimation register is in
the state $\ket{\psi'}$. If the estimation register was in the state 
$\ket{\psi}$, uncomputing the eigenvalue estimation would lead to the correct
initial state $\ket{0}$. If it is in the state $\ket{\psi'}$, then, after uncomputing
the eigenvalue estimation, $E$ can be in a basis state different from $\ket{0}$
with probability at most $\|\psi-\psi'\|^2 = \|\psi''\|^2 \leq \epsilon$.

Thus, the probability of the computation terminating for a fixed $j<j_i$
is at most $2\epsilon$. The probability of that happening for some $j<j_i$
is at most $2(j_i-1) \epsilon<2(m-1)\epsilon$.
\qed

We now assume that the algorithm is started from stage $j_i$.

\begin{Claim}
\label{cl:c2}
If Algorithm \ref{alg:amplify} is started from stage $j_i$ (instead of stage 1), then 
\[ \left\|  \ket{\phi_{i, j_i}}_E - \frac{1}{\kappa \tilde{\lambda}} \ket{0}_E \right\|^2 
\leq \epsilon \left( 1 + \frac{3\epsilon}{2} \right) C 
.\]
\end{Claim}

\proof
Let
\[\ket{\psi} = \sum_{\lambda} \alpha_{\lambda} \ket{\lambda} \] be the
output of the eigenvalue estimation in stage $j_i$. Then, 
$|\alpha_{\tilde{\lambda}_i}|^2 \geq 1-\epsilon$ and
$\| \ket{\psi} - \alpha_{\tilde{\lambda}_i} \ket{\tilde{\lambda_i}}\|^2 \leq \epsilon$.
Conditional on $O$ being mapped to $\ket{1}$, 
the estimation register $E$ is in the state
\[ \ket{\psi'} = \sum_{\lambda} \beta_{\lambda} \ket{\lambda} \]
where $\beta_{\lambda}=\frac{\alpha_{\lambda}}{\kappa \lambda}$ when
$\lambda\geq \frac{1}{\epsilon 2^{j+1}}$ and $\beta_{\lambda}=0$ otherwise.
By Claim \ref{cl:doubling}, we have
\[ \frac{1}{\lambda} \in [0, \epsilon 2^{j+1}]
\subseteq  \left[0, \left( 2 + \frac{3\epsilon}{2} \right) \frac{1}{\tilde{\lambda}}\right] .\]
When
$\lambda\geq \frac{1}{\epsilon 2^{j+1}}$, this implies 
\[ \left| \beta_{\lambda} - 
 \frac{\alpha_{\lambda}}{\kappa\tilde{\lambda} } \right| =
\left| \frac{\alpha_{\lambda}}{\kappa\lambda} - 
 \frac{\alpha_{\lambda}}{\kappa\tilde{\lambda} } \right|
\leq  \left( 1 + \frac{3\epsilon}{2} \right)
\frac{\alpha_{\lambda}}{\kappa\tilde{\lambda} } .\]
When $\lambda< \frac{1}{\epsilon 2^{j+1}}$, we have $\beta_{\lambda}=0$ and
\[ \left| \beta_{\lambda} - 
 \frac{\alpha_{\lambda}}{\kappa\tilde{\lambda} } \right| =
\frac{\alpha_{\lambda}}{\kappa\tilde{\lambda} } .\]
By summing over all $\lambda\neq \tilde{\lambda}$, we get
\[ \left\| \psi' - \frac{1}{\kappa \tilde{\lambda}} \psi \right\|^2 \leq 
\left( 1 + \frac{3\epsilon}{2} \right) C \sum_{\lambda: \lambda\neq \tilde{\lambda}} 
|\alpha_{\lambda}|^2 \leq
\left( 1 + \frac{3\epsilon}{2} \right) \epsilon C
 .\]
Therefore, (conditional on the outcome register being 
$\ket{1}$) uncomputing {\bf UniqueEst} leads to a state 
$\ket{\varphi}_E$ with 
\[ \left\| \varphi - \frac{1}{\kappa \tilde{\lambda}} \ket{0} \right\|^2 
\leq \epsilon \left( 1 + \frac{3\epsilon}{2} \right) C .\]
\qed

Since the algorithm might not reach stage $j_i$ with probability at most 
$2(m-1)\epsilon$, we have to combine the error bounds from Claims \ref{cl:c1}
and \ref{cl:c2}. This gives us
\[ \left\|  \ket{\phi_{i, j_i}}_E - \frac{1}{\kappa \tilde{\lambda}} \ket{0}_E \right\| 
\leq \epsilon \left( 2m - 1 + \frac{3\epsilon}{2} \right) C 
.\]

Combining this with bounds of $1.6\epsilon C$ and $36\epsilon C$ on
$\|\psi_{<}\|$ and $\|\psi_{>}\|$ completes the proof of Claim \ref{cl:good}.
\qed

\begin{Claim}
\label{cl:bad}
If $\ket{v_i}$ is bad, 
\[ \| \phi'_i \|^2 \leq 30 C\]
where $C=(\frac{1}{\kappa}{\tilde{\lambda}_i})^2$. 
\end{Claim}

\proof
We express
\[ \ket{\phi'_i} = \ket{\phi_{\leq}}+\ket{\phi_{>}} \] 
where
\[ \ket{\phi_{\leq}}=\sum_{j\leq j_i+1} \ket{2j}_S \otimes \ket{\phi_{i, j}}_E, \]
\[ \ket{\phi_{>}}=\sum_{j>j_i+1} \ket{2j}_S \otimes \ket{\phi_{i, j}}_E.\]
We have 
\begin{equation}
\label{eq:bad1} 
\|\phi_{\leq} \|^2 \leq \left(\frac{1}{\kappa \epsilon 2^{j_i+2}}\right)^2 \leq 
16 \left(1+\frac{3\epsilon}{2} \right)^2 C .
\end{equation}
Here, the first inequality follows from the amplitude of $\ket{1}$ in (\ref{eq:transform})
being $\frac{1}{\kappa \lambda}$, $\lambda\geq \frac{1}{\epsilon 2^{j+1}}$
and $j\leq j_i+1$. The second inequality follows from Claim \ref{cl:doubling}.

Starting from stage $j+1$, the probability of algorithm obtaining
$\lambda<\frac{1}{2^{j+1}\epsilon}$ is at most $\epsilon$ at each stage. 
Therefore (similarly to the proof of Claim \ref{cl:good}),
\[ \|\phi_{>}\|^2 = \sum_{j>j_i+1} \|\phi'_{i, j}\|^2 \leq
\sum_{j>j_i+1} \left( \frac{2^{j+1} \epsilon}{\kappa} \right)^2 \epsilon^{j-j_i-1} \leq
\left( \frac{2^{j_i+2}\epsilon}{\kappa} \right)^2 \sum_{j=1}^{\infty} (4\epsilon)^j  \leq \]
\begin{equation}
\label{eq:bad2} 
 16 \left(1+\frac{3\epsilon}{2} \right)^2 C 
\sum_{j=1}^{\infty} (4\epsilon)^j =
16 \left(1+\frac{3\epsilon}{2} \right)^2 \frac{4\epsilon}{1-4\epsilon} C .
\end{equation}
The claim follows by putting equations (\ref{eq:bad1}) and (\ref{eq:bad2}) together and using $\epsilon<0.01$.
\qed

\proof [of Lemma \ref{lem:close1}]
We have
\[ |\lambda_i - \tilde{\lambda}_i | \leq \frac{1+\epsilon}{2^{j+1}} \leq (1+\epsilon) \epsilon \tilde{\lambda}_i,\]
with the first inequality following from the correctness of the unique-output eigenvalue estimation and
the second inequality following from the definition of $\tilde{\lambda}_i$.
Let $\delta=(1+\epsilon)\epsilon$.

If $|\lambda_i-\tilde{\lambda_i}| \leq \delta \tilde{\lambda_i}$, then 
\[ \left|\frac{1}{\lambda_i}-\frac{1}{\tilde{\lambda_i}} \right| \leq \frac{\delta}{1-\delta}  
\frac{1}{\tilde{\lambda_i}} .\]
Therefore, we have $\| \ket{\psi'} - \ket{\psi_{ideal}} \| \leq \frac{\delta}{1-\delta} \| \ket{\psi_{ideal}} \|$ and
\[ \frac{\delta}{1-\delta} = \frac{(1+\epsilon)\epsilon}{1-(1+\epsilon)\epsilon} < \frac{2\epsilon}{1-2\epsilon} .\]
\qed

\subsection{Proofs of Lemmas about the running time of Algorithm 4}
\label{sec:proofs1}

\proof [of Lemma \ref{lem:l2}]
We first consider the case when the input state $\ket{x}$ is an eigenstate
$\ket{v_i}$ of $H$.
Let $p_{stop, j}$ be the probability that Algorithm \ref{alg:amplify} stops after stage $j$.
Then, the square of $l_2$ average running time of Algorithm \ref{alg:amplify} is of the order
\begin{equation}
\label{eq:lem2a} 
O\left( \sum_j p_{stop, j} 2^{2j} k^2_{uniq} \right)
\end{equation}
since, in first $j$ stages we use amplitude amplification for time
\[ k_{uniq} (2+2^2+\ldots +2^j) = k_{uniq} (2^{j+1}-2) = O(k_{uniq} 2^j) .\] 
Let $j\geq j_i+1$.
The probability that, in the $j^{\rm th}$ run of eigenvalue estimation,
the algorithm does not stop is at most $\epsilon$.
Therefore, $p_{j_i+k}\leq \epsilon^{k-1}$ and the expression in (\ref{eq:lem2a}) is at most $k^2_{uniq}$ times
\[  2^{2(j_i+1)} + \sum_{j=j_i+2}^{\infty} 
\epsilon^{j-j_i-1} 2^{2j} 
< 2^{2(j_i+1)} + 2^{2(j_i+1)} \sum_{j=1}^{\infty} (4\epsilon)^j 
= O(2^{2j_i})  .\]

If $\ket{x}=\sum_i \alpha_i \ket{v_i}$, the square of $l_2$-average of the number
of steps is of the order
\[  O\left(\sum_{i} |\alpha_i|^2 2^{2 j_i} k_{uniq}^2 \right) \]
because, each subspace of the form $\ket{v_i}\otimes \H_A \otimes \H_S \otimes \H_E$
stays invariant throughout the algorithm and, thus, can be treated separately.
Taking square root finishes the proof.
\qed

\proof [of Lemma \ref{lem:prob}]
Again, we can treat each subspace of the form 
$\ket{v_i}\otimes \H_A \otimes \H_S \otimes \H_E$ separately.
As shown in the proof of Claim \ref{cl:good},
the probability of the algorithm stopping before stage $j_i$ is at most
$2(j_i-1) \epsilon\leq 2(m-1) \epsilon$. 
Therefore, the algorithm stops at stage $j_i$ or $j_i+1$ with 
a probability that is at least a constant.
The probability of algorithm stopping succesfully (i.e., producing $\ket{1}$ in
an outcome register) is $\frac{1}{\kappa^2 \lambda^2}$.
By Claim \ref{cl:doubling}, we have $\lambda=O(\frac{1}{\epsilon 2^{j_i}})$.
This implies that the probability of the algorithm stopping successfully
is $O(\frac{\epsilon^2 2^{2j_i}}{\kappa^2})$.
\qed

\end{document}